\documentclass{sigchi}

\CopyrightYear{2016}
\setcopyright{acmlicensed}
\doi{http://dx.doi.org/10.475/123_4}
\isbn{123-4567-24-567/08/06}
\conferenceinfo{CHI'16,}{May 07--12, 2016, San Jose, CA, USA}
\acmPrice{\$15.00}

 \toappear{
Permission to make digital or hard copies of all or part of this work for personal or classroom use is granted without fee provided that copies are not made or distributed for profit or commercial advantage and that copies bear this notice and the full citation on the first page. To copy otherwise, or republish, to post on servers or to redistribute to lists, requires prior specific permission and/or a fee. IHC'17, Brazilian Symposium on Human Factors in Computing Systems. October 23-27, 2017, Joinville, SC, Brazil. Copyright 2017 SBC. ISBN XXX-XX-XXXX-XXX-X (pendrive).

 }

\usepackage{balance}       
\usepackage{graphics}      
\usepackage[T1]{fontenc}   
\usepackage{txfonts}
\usepackage{mathptmx}
\usepackage[pdflang={en-US},pdftex]{hyperref}
\usepackage{color}
\usepackage{booktabs}
\usepackage{textcomp}

\usepackage{cite}
\usepackage{subfigure}
\usepackage{graphicx}
\usepackage[flushleft]{threeparttable}

\usepackage{microtype}        
\usepackage{ccicons}          

\usepackage{todonotes}

\def\plaintitle{Designing for Pragmatists and Fundamentalists: Privacy Concerns and Attitudes on the Internet of Things}

\def\emptyauthor{}
\def\plainkeywords{Privacy perceptions, Concerns about privacy, Internet of Things, Information Boundary, Face Keeping.}

\makeatletter
\def\url@leostyle{%
  \@ifundefined{selectfont}{
    \def\UrlFont{\sf}
  }{
    \def\UrlFont{\small\bf\ttfamily}
  }}
\makeatother
\def\pprw{8.5in}
\def\pprh{11in}

\definecolor{linkColor}{RGB}{6,125,233}
\hypersetup{%
  pdftitle={\plaintitle},
  pdfauthor={\emptyauthor},
  pdfkeywords={\plainkeywords},
  pdfdisplaydoctitle=true, 
  bookmarksnumbered,
  pdfstartview={FitH},
  colorlinks,
  citecolor=black,
  filecolor=black,
  linkcolor=black,
  urlcolor=linkColor,
  breaklinks=true,
  hypertexnames=false
}

\begin{document}

\title{\plaintitle}

\numberofauthors{3}
\author{%
  \alignauthor{Lesandro Ponciano\\
    \affaddr{Pontifical Catholic University of Minas Gerais}\\
    \affaddr{Belo Horizonte, Brazil}\\
    \email{lesandrop@pucminas.br}}\\
  \alignauthor{Pedro Barbosa\\
    \affaddr{Federal University of Campina Grande}\\
    \affaddr{Campina Grande, Brazil}\\
    \email{pedroyossis@copin.ufcg.edu.br}}\\
  \alignauthor{Francisco Brasileiro\\
    \affaddr{Federal University of Campina Grande}\\
    \affaddr{Campina Grande, Brazil}\\
    \email{fubica@dsc.ufcg.edu.br}}\\
  \alignauthor{Andrey Brito\\
    \affaddr{Federal University of Campina Grande}\\
    \affaddr{Campina Grande, Brazil}\\
    \email{andrey@computacao.ufcg.edu.br}}\\
  \alignauthor{Nazareno Andrade\\
    \affaddr{Federal University of Campina Grande}\\
    \affaddr{Campina Grande, Brazil}\\
    \email{nazareno@computacao.ufcg.edu.br}}\\
}

\maketitle

\begin{abstract}
Internet of Things (IoT) systems have aroused enthusiasm and concerns. Enthusiasm comes from their utilities in people daily life, and concerns may be associated with privacy issues. By using two IoT systems as case-studies, we examine users' privacy beliefs, concerns and attitudes. We focus on four major dimensions: the collection of personal data, the inference of new information, the exchange of information to third parties, and the risk-utility trade-off posed by the features. Altogether, 113 Brazilian individuals answered a survey about such dimensions. Although their perceptions seem to be dependent on the context, there are recurrent patterns. Our results suggest that IoT users can be classified into unconcerned, fundamentalists and pragmatists. Most of them exhibit a pragmatist profile and believe in privacy as a right guaranteed by law. One of the most privacy concerning aspect in IoT is the exchange of personal information to third parties. Also, individuals' perceived risk tends to be negatively correlated with their perceived utility in the features of the system. We discuss practical implications of these results and suggest heuristics to cope with privacy concerns when designing IoT systems.
\end{abstract}
  


\section{Introduction}
\label{sec:introduction}

The Internet of Things (IoT) is composed of devices that people use in their daily life. Besides the traditional desktops, many other devices like vehicles, buildings, and home appliances are on the network~\cite{gubbi2013internet}. The devices are embedded with network connectivity that enable them to collect and exchange data~\cite{Elkhodr2012}. Systems based on such network can aggregate data collected by various devices. The collected data may allow the system to devise users' a unique identifier, and monitor data as where they have been, with whom they have been and what actions they have taken. Based on machine learning and data mining algorithms, system can infer information that were not directly collected from the users and predict their behaviors, preferences and choices~\cite{lee2011,Shokri2011}. Such information can be used to provide features to them. This sort of systems has aroused enthusiasm and concerns. Enthusiasm comes from their potential utilities in people daily life, and concerns are mainly associated to their potential threats to people's privacy.

There is no simple and widely accepted definition of privacy~\cite{Schwaig2013,baruh2014,yao2007}. Different notions of privacy have been object of analysis in studies from various disciplines such as psychology, sociology and computer science. Even in computer science, privacy has been approached from a variety of perspectives depending on the type of system, such as online social networks~\cite{Rodrigues:2016,barnes2006privacy,Lopes:2016}, mobile apps~\cite{Yamauchi:2016,dai2007,king2010,Elkhodr2012}, healthcare applications~\cite{tentori2006,Chen2012}, and so on. In the context of information systems, one definition of privacy is the individual's control over the collection, disclosure, and use of her/his personal information~\cite{Alhalafi2015,dai2007,Mekovec2011}. If such control does not exist or is compromised, the privacy of the individuals is not being guaranteed~\cite{dai2007,Mekovec2011,Zhou2015}. When they perceive that the system invades their privacy or that is not developed to prevent violations from third parties, users associate a risk to the use of the system. Faced with this risk, users may decide do not engage in the system or restrict their level of engagement~\cite{Li2014,dai2007,Nov2009}. To successfully take into account the privacy of their users, the design of IoT systems must adequately cope with users' concern about privacy.

Naturally, different people may not perceive privacy and react to issues in the same way. Individuals can be broadly categorized into profiles according to their levels of privacy concerns, their attitudes, and what the system can do using their information~\cite{sheehan2002,yau2007,lee2011,baruh2014}. One widely used categorization is based on three privacy attitude profiles: privacy unconcerned users; privacy pragmatist users; and the privacy fundamentalist users~\cite{yau2007,lee2011}. The unconcerned privacy attitude profile consists of individuals who are inclined to share information about them, regardless of actions taken by the system. Individuals who exhibit a pragmatist privacy attitude profile condition the data provision to action taken by the system, as adopting a privacy protection mechanism. Finally, individuals who exhibit a fundamentalist privacy attitude profile are not inclined to share information about them. Little is known so far about the occurrence of these profiles in IoT systems and how to develop IoT systems that meet their privacy needs.

We sought to deeply investigate people's privacy beliefs, concerns, and attitudes toward IoT systems. To do so, we address the following research questions: 1) what is the occurrence and characteristics of fundamentalist, pragmatic and unconcerned users in IoT systems; 2) which components of IoT systems can cause more concern in terms of privacy; 3) how do users perceive the risk-utility trade-off posed by features of IoT systems. To answer these questions, we surveyed 113 potential users of 2 IoT systems. The study was conducted with students and technology professionals in Brazil. We classified the users into the three privacy attitude profiles: unconcerned, fundamentalists and pragmatists. Based on the classification, we analyze the privacy beliefs, concerns, and perceptions.

Although several of the results seem to be dependent on the context of each system, there are some patterns across systems that can be highlighted. We find that most of the individuals exhibit a pragmatist privacy attitude profile. They also tend to believe in privacy as a right guaranteed by law. Regardless the privacy profile, the most privacy concerning aspect of IoT systems tend to be the exchange of information to third parties, showing average levels of concerns higher than those of data collection and information inference. About the risk-utility trade-offs posed by the systems, the perceived risk tends to be negatively associated with their perceived utility. The higher the perception of utility in the feature of the system, the lower the perceived risk in such feature. 

We discuss several practical implications of these results and suggest heuristics to cope with privacy concerns in the design of IoT systems. In summary, we suggest that system designers 1) let users know what information the system has about them; 2) make clear the usefulness of the data for each feature; 3) make the exchange of data with third parties configurable; 4) conduct empirical assessments of privacy to deal with cases that are system dependent.

The rest of this work is organized as follows. We provide first a background of key concepts related to privacy and discuss relevant previous work. Next we discuss our approach to characterize privacy beliefs, concerns, and attitudes. It is followed by the evaluation study in two IoT Systems; we discuss the implications of the results and propose heuristics to cope with privacy concerns in the design of IoT systems. Finally, we summarize the conclusions of the study.

\section{Background and Related Work}
\label{sec:background}

In this section, we first briefly review relevant concepts of privacy. After that, we detail two theoretical constructs to explain people's privacy concerns and attitudes. Finally, we analyze what is known about privacy concerns and attitudes in the context of IoT systems.

\subsection{What is privacy?}

From a legislation viewpoint, considering the laws in the United States of America, privacy can be defined as ``the right of an individual to be let alone''~\cite{warren1890right}. People, in turn, usually associate the word privacy with a diversity of meanings. Some people believe that privacy is the right to control what information about them may be made public~\cite{Schwaig2013,baruh2014,yao2007}. Other people believe that if someone cares about privacy is because he/she is involved in wrongdoing~\cite{Beckwith2003}. Privacy is also associated with the states of solitude, intimacy, anonymity, and reserve~\cite{lahlou2008,yao2007}. Solitude means the physical separation from other individuals. Intimacy is some kind of close relationship between individuals with which information is exchanged. Anonymity is the state of freedom from identification and surveillance. Finally, reserve means the creation of psychological protection against intrusion by other unwanted individuals. 

In information and communications technology (ICT), the concept of privacy is usually associated to the degree of control over the flow of personal information~\cite{dai2007}. In this context, people associate privacy to something regarding to their level of control over the collection of personal information, and usage of the collected information, and the third parties that can have access to the information, such as relatives, friends, hierarchical superiors, and government agencies~\cite{Alhalafi2015,dai2007,Mekovec2011}.

\subsection{Do people care about privacy?}

Concerns about privacy usually arise from unauthorized collection of personal data, unauthorized secondary use of the data, errors in personal data, and improper access to personal data~\cite{smith1996}. People concerns are indeed associated to possible consequences that these occurrences may have on their lives. Two relevant theoretical constructs that explore this view are face keeping and information boundary.

\subsubsection{Face keeping}

Face keeping considers that to perform social interaction successfully, people must provide the others with some identity~\cite{lahlou2008}. Providing some identity is to share information, such as providing information about the tastes, preferences, beliefs, and past actions. In doing so, individuals construct a public ``face'' of themselves made with the features that they are sharing. In daily life, people engage in a variety of activities. They do not necessarily use the same ``face'' in all activities. For example, in some instances they may want to play the role of a friendly individual, while in other they may want to play the role of an individual who is expert in a specific domain. What people would say, do and disclose when wearing these different faces may be quite different. In general, ``faces'' are a combination of roles and status. 

People care about keeping a public face that pleases them. Thus, from a ``face keeping'' perspective, a privacy breach is a case of losing a face, in which people are forced to put on or endorse an unwanted face. It occurs, for example, when some people want to maintain their control over the information about their conditions or past, such as people who carry certain diseases such as AIDS, ex-convicts, or members of some religious groups. To avoid stigma, they seek to prevent their information from being disseminated and become part of their faces. The privacy concerns come from the need to maintain a desired public face.

\subsubsection{Information boundary}

Information boundary explains the dynamics of individual privacy concern based on the metaphor of two inner states: ``boundary opening'' and ``boundary closure''~\cite{xu2011}. When the boundary is open, the individual reveals information freely. When the boundary is closed, the information flow is restricted. Individuals establish privacy rules to define when the boundary is open and when it is closed. Each individual has a mental calculus that is used to construct rules based on the risk-benefit calculation of information disclosure~\cite{dinev2006extended}.

After individuals disclose their personal information to somebody or something, the information moves to a collective domain where all the co-owners of such information share a joint responsibility for keeping it private. The main concern with privacy is that information leaks beyond this domain. Privacy rules used in the decision to disclose personal information are based on the expectations about who will have access to the information in the collective domain. Policies about the protection of the information in collective domain are negotiated before the information is disclosed. When the policies are not met, individual boundary management becomes turbulent and the individual becomes more restrictive in their privacy rules, recalculating their risk-benefit in information disclosure.

\subsection{What is known about individual privacy concerns and attitudes toward IoT systems?}

Studies have reported a number of aspects of IoT systems that may cause privacy concern, among them we highlight:
\begin{itemize}
\item The potential to enable a systematic mass surveillance and to impinge on the personal privacy of users, especially their location privacy~\cite{Elkhodr2012}.
\item The automatic collection of data from the network of devices and their secure protocols cannot be verified by the data owners~\cite{wong2014towards}.
\item The increase of privacy concerns because of three main factors: they are based on novel, heterogeneous and constrained user interfaces; the ubiquitous presence of devices and their potential to vast data collection; and market forces and misaligned incentives~\cite{Williams2016}
\end{itemize}

About users' privacy attitudes, some studies discuss that in face privacy concerns people my refuse to provide their personal data or submit inaccurate data~\cite{wong2014towards}. Other studies, however, identify that privacy concerns have a relatively weak effect on influencing the adoption of IoT systems~\cite{Hsu2016}, being other factors more relevant in terms of adoption. In fact, in the literature on people's privacy perception, the association between privacy concerns and privacy attitudes is usually confusing and paradoxical. This phenomenon is known as the {\it privacy paradox}, in which people claim to care about privacy, but they are often perceived to act to the contrary~\cite{barnes2006privacy,norberg2007privacy}. It is believed that this paradox must also occur in IoT~\cite{Williams2016}.

\section{Characterizing privacy beliefs, concerns and attitudes}
\label{sec:approach}

Our approach to characterize users' privacy concerns and attitudes toward IoT systems consists of three dimensions: (i) demographics, general privacy beliefs, attitudes; (ii) sources of privacy concerns in IoT system; and (ii) risk-utility trade-offs posed by the features of the system. Throughout this section we discuss how we address each of these dimensions.
 
\subsection{Demographics, general privacy beliefs and attitudes}

This dimension focuses more on individual general characteristics than on characteristics they exhibit when using a specific system. We consider a set of characteristics that may be insightful about how do users perceive privacy issues in IoT systems, which are their demographics, beliefs, and attitudes.

In characterizing demographics characteristics, we consider gender, education, and age. In designing a survey instrument, we ask respondent to inform their gender (females, male or inform other) and education (basic school, high school, undergraduate, master and doctoral degrees) and age (Under 18 years old, 18--24 years old, 25--34 years old, 35--44 years old, 45--54 years old, 55 years or older). 

In assessing individual beliefs, we consider a set of privacy beliefs reported in previous studies of privacy perceptions, which are: privacy as a right guaranteed by law, privacy as an individual's responsibility, and privacy as a need associated to people who are involved in wrongdoing. We asked users to select the option that best describes their feelings, which are: (a) {\it I believe that privacy is a right guaranteed by law}, (b) {\it I believe that each person has a responsibility to protect his/her own privacy}, and (c) {\it I believe that people who cares about privacy is because he/she is involved in wrongdoing}. These privacy beliefs may be insightful about how do users perceive privacy issues in IoT systems. 

Finally, privacy attitudes are investigated by asking users about how they are likely to act when they are requested to provide private data to a system. We provide to the respondents the following instruction and ask them to select the option that is closest to the way they see the situation: {\it If you are using a system and it asks you to provide information that you consider personal, which of these behaviors you adopt? (a) You provide the requested information. (b) You provide the requested information only if the system informs its privacy policy. (c) You provide the requested information only if you know that the system will give you something in return. (d) You do not provide the requested information.} It is the core point in the categorization of users in the three profiles: unconcerned; pragmatist; and fundamentalist~\cite{yau2007,lee2011}. The chosen option allows to infer user's privacy attitude profile: unconcerned (answer {\it a}), fundamentalist (answer {\it d}), and pragmatist (answer {\it b} or {\it c}).

\subsection{Sources of privacy concerns in IoT systems}

Three components of typical IoT systems that can be source of users' privacy concerns are: the collection of personal data from users; the inference of richer information based on the collected data; and the exchange of users' information with third parties.

\paragraph{Data collection} Systems can collect data directly from users by asking them to provide data or indirectly via their devices, such as smart phones and smart watches. In order to investigate the level of privacy concern caused by each collected data, we ask users to inform the level of privacy concern in a 5-point Likert scale. For instance, about location information collected by mobiles, we can ask: {\it Assuming that the system is able to collect the locations where your mobile has been throughout the day, how concerned would you feel about it? (a) Not concerned at all; (b) Slightly concerned; (c) Moderately concerned; (d) Very concerned; (e) Extremely concerned.}. We also ask for the level of concerns caused by the collection of information about choices made by the user, and information about other users with who the user meets.

\paragraph{Inference of richer information} The data collected by the system can be used to find new information about the users. Systems can perform information inference for a variety of purposes, such as identifying users' friends, and recommending products and services to them. In order to investigate the level of privacy concern caused by each piece of information inferred by a system, we ask users to inform the level of privacy concern in a 5-point Likert scale. For instance, about the inference of favorite TV shows, we can ask: {\it Assuming that when using the system it would be able to infers your favorite TV shows, how concerned would you feel about it? (a) Not concerned at all; (b) Slightly concerned; (c) Moderately concerned; (d) Very concerned; (e) Extremely concerned.}

\paragraph{Exchange of information with third parties} As part of its operation or due to a failure of operation, system can let users' data be accessible to third parties. Users consider this kind of situation as an improper access to their personal data~\cite{lahlou2008}. To whom information is made accessible is relevant, for example, users' friends, relatives, employers, government agencies, and other systems~\cite{Kwasny2008,steijn2015}. In order to investigate the level of privacy concern caused by information exchange, for each piece of information handled by the system, we ask users to inform the level of privacy concern caused by the data being provided to third parties in a 5-point Likert scale. For instance, about the data being provided to government agencies, we can ask: {\it Enter the level of concern you feel if the data about the places where your mobile was throughout the day are provided to government agencies: (a) Not concerned at all; (b) Slightly concerned; (c) Moderately concerned; (d) Very concerned; (e) Extremely concerned}.

All the questions about the sources of privacy concerns (data collection, inference of richer information, and exchange of information with third parties) are based on in a 5-point Likert scale. It is an ordinal scale, so we can say for example that the "Extremely concerned" score is higher than the ``Very concerned'' score. When analyzing the answers, we assign numbers to the scale, so 1 is lower value, 3 is the neutral response, and 5 is the higher value. We assess the level of consensus in the answers provided by the users~\cite{ponciano2014considering,krippendorff1970estimating}. To measure the level of consensus we use the Krippendorff's alpha coefficient~\cite{krippendorff1970estimating}. The maximum value to this metric is 1, indicating complete consensus among the respondents, and the minimum is -1, indicating complete lack of consensus.

In this characterization of the sources of privacy concerns in IoT system, we are usually interested in identify if the average of users privacy concerns lies below or above the neutral concern (3 -- ``Moderately concerned''). In this case, the neutral concern serves as an arbitrary zero point.

\subsection{System information use trade-offs} 

In order to evaluate the risk-benefit trade-off of features in IoT systems, we also measure the level of utility perceived by users in the feature that use such information. To do so, we ask users to inform the level of utility of the feature in a 5-point Likert scale. For example, in the case of TV shows schedule features, we can ask users {\it How useful do you consider the feature that infers your favorite TV shows and automatically turns on the TV on their schedule? (a) Not useful at all; (b) Slightly useful; (c) Moderately useful; (d) Very useful; (e) Extremely useful.}

Based on the measures of privacy concern caused by the data collection/inference and the measures of utility of the feature that use the data, we measure the degree of correlation between the risk indicated by the concern and utility reported by the users. When there is a correlation, we shape it into an equation by means of a regression. Thus, the trade-off can be evaluated as the model of how an increase in utility of a feature impacts the risk perceived by the users.

\section{Privacy in Two IoT Systems}
\label{sec:evaluation}

By using the approach discussed in the previous section, we characterize users' privacy beliefs, concerns and attitudes in two IoT systems: Lumen and Pulso. We begin this section by introducing these systems. After that, we detail our survey, the recruitment of participants, and the results of our study.

\textbf{Pulso system} measures the ``pulse'' of locations, i.e. how many people are there. It is based on people counting sensors that are installed in specific locations, such as restaurants, public squares, and bus stop points. The sensors work just like internet access points. When users arrive at the location with their smart phones with the WiFi turned on, the smart phones probe for connections. A smart phone probing connections is considered as a person in the location; the smart phones are uniquely identified by their Media Access Control address (MAC address). The number of people in a location is estimated by the number of smart phones at the location that have WiFi on. Pulso system collects and stores the time and MAC address of each smart phone probing a connection. Such data is processed in order to provide features/services to their users. The features are accessible to the users via a mobile application. In the application, users can search for locations and, for each searched location, the system reports in real time how many people are there. Users can use it for different things. For instance, they can use it to know how many people are in the restaurant they want to lunch, and decide to go in a less crowded time.

\textbf{Lumen system} provides notifications of disproportionate power consumption in an organization. The system combines information of the number of people with the energy consumption. The number of people is counted by Lumen in the same way that Pulso does. Sensors are installed in the organization. When people arrive with their smart phones with the WiFi connection turned on, the smart phone probes for WiFi connections. The number of people is the number of unique MAC addresses. The energy consumption in the organization is obtained by using a smart power meter. The collected data is processed and presented to users via a Web application. The application allows users to check the environmental impact history -- the ratio between the energy consumption and the number of people in the organization. It informs the moments in time when there is an above-average environmental impact. At such times, the system generates a notification that alerts users about power consumption that is disproportionate to the number of people in the organization. The notifications are made available in a notification board that can be visualized by everyone in the organization.

\textbf{Relevance of Pulso and Lumen to study privacy in IoT:} Both systems involve the face keeping concept. In Pulso, one of users' privacy concerns may be that information about places where they go routinely becomes public. In Lumen, the concern is that the system publicly assigns to users the responsibility for an energy consumption problem in the organization. Thus, in both cases, the system can potentially change the public face that the users wear. The two systems also involve information boundary concept. In both cases, people seek to control what information about them may belong to the public domain. Also, the systems involve different spaces and forms of adoption. Pulso is designed for a public space and anyone can join and leave the system at anytime. Lumen, in turn, is designed for a corporate space and members of the organization are joined to the system as part of their activities.

\textbf{Survey design and recruitment of participants:} To assess individuals privacy perceptions and concerns towards the Pulso and the Lumen system, we build a survey with a set of questions derived from the proposed approach (Section~\ref{sec:approach}). The dynamic of the system is explained to the respondent by presenting screens of the application and a video prototype that explain how the system works. The survey was designed by using Google Forms tools\footnote{The forms are currently retired and permanently available at https://goo.gl/forms/2HpdQmvg9uHCL86f1
(Pulso) and https://goo.gl/forms/y9W1TVrxQ81mPI483
(Lumen).}.

After validating the survey in pilot tests, we collected sample of answers from potential users of the systems. For the Pulso systems, answers were collected from 58 respondents in a transportation hub in Campina Grande, Brazil. For the Lumen system, answers were collected from 55 participants that work in computer science research labs and in a software development company in Campina Grande, Brazil. In both studies, respondents answered the survey individually and without any external interference.

\subsection{Results}

As summarized by our research questions, our main objective is to characterize individuals privacy concerns and attitudes toward IoT systems. The results of the characterization are organized in three parts presented in the following. We first analyze users' demographics and privacy beliefs according to their privacy attitude profiles. Next we discuss the sources of their privacy concerns in IoT systems (collection, inference and exchange of information). Finally, we present the results on individuals perceptions of the risk-benefit trade-offs.

\subsubsection{Privacy profiles: distribution, demographics and consensus}

\begin{figure*}[htb]
\centering
\subfigure[Pulso]{
\includegraphics[width=.47\linewidth]{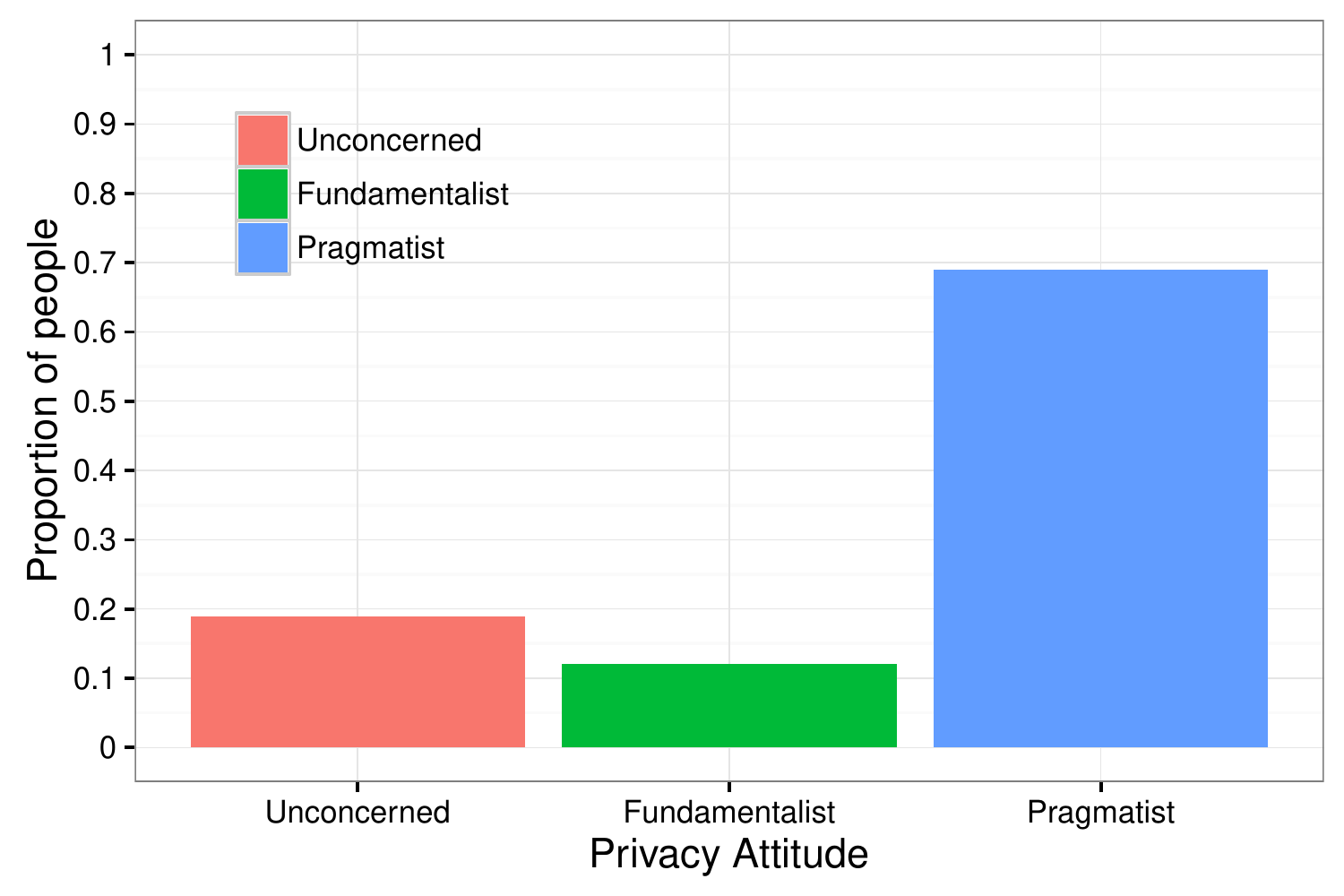}
\label{subfig:pulso:profilesdistribution}
}
\subfigure[Lumen]{
\includegraphics[width=.47\linewidth]{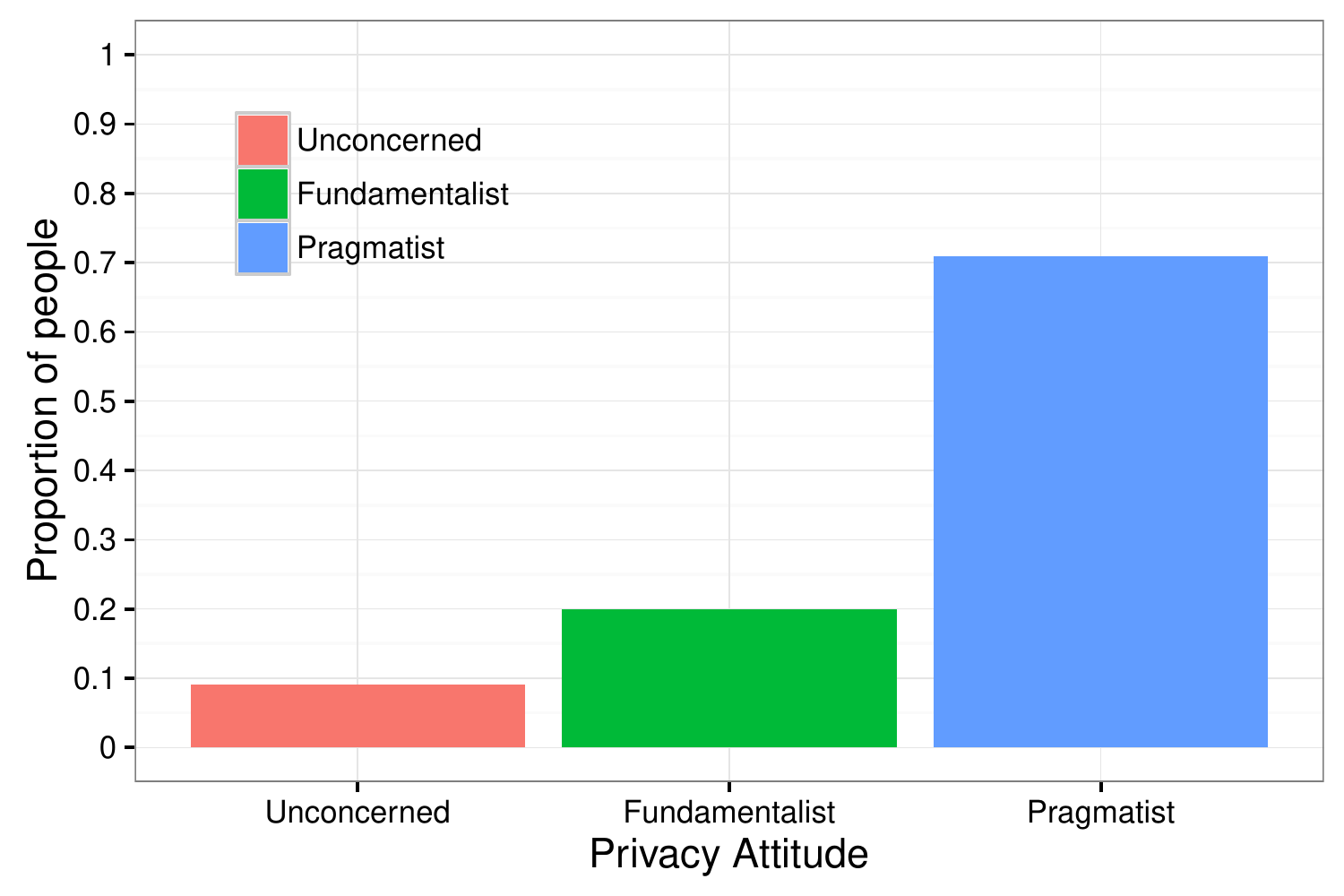}
\label{subfig:lumen:profilesdistribution}
}
\vspace{-8pt}
\caption{The proportion of respondents who falls into each privacy attitude profile: unconcerned, pragmatists, and fundamentalists.}
\label{fig:profilesdistribution}
\end{figure*}

We identified people who exhibited the three privacy attitude profiles: unconcerned, fundamentalists, and pragmatist. As Figure~\ref{fig:profilesdistribution} shows, the pragmatist is the most frequent profile. Most respondents exhibits this privacy attitude profile in the Pulso system (40 individuals or a proportion of 0.69) and in the Lumen system (39 individuals or a proportion of 0.71).

Table~\ref{table:demographicsPulso} and Table~\ref{table:demographicsLumen} show the statistic summary of the demographic characteristics and privacy beliefs of participants in the Pulso and Lumen systems, respectively. These statistics show that demographics and beliefs of participants who fall in each privacy attitude profile change according to the set of respondents who participated in the analysis of each system. We summarize profiles demographics and beliefs as follows:

\begin{itemize}
\item {\bf Unconcerned.} Most of the users who exhibit an unconcerned privacy attitude profile in the Pulso system are female with age ranging from 18 and 24 years old and that completed just high school. In the Lumen System, all of them are male respondents, and most of them have age ranging from 25 and 34 years old and completed undergraduate degree. The unconcerned users are practically divided between those who believe in privacy as a right and those who believe in privacy as a personal responsibility of each user.
\item {\bf Fundamentalists.} In the Pulso system, most of them are female respondent with age ranging from 18 and 24 years old and that completed just high school. In the Lumen system, most of them are male respondent with age ranging from 25 and 34 years old and that completed undergraduate degree. Users who exhibit a fundamentalist are divided into those who believe in privacy as a right and those who believe in privacy as a personal responsibility.
\item {\bf Pragmatists.} Most of them are female respondents with age ranging from 18 and 24 years old and that completed undergraduate degree. In the Lumen system, the majority of fundamentalists are male respondents, with age ranging from 25 and 34 years old and that completed undergraduate degree. In both systems, the pragmatist users are more likely to believe in privacy as a right.
\end{itemize}

\begin{table}[ht]
    \small
    \begin{tabular}{p{2.503cm}|p{1.11cm}|p{1.11cm}|p{1.0cm}|p{1.0cm}}
    \hline
        & Unconcern. & Fundament. &  Pragmat. & {\it}All \\ \hline
\textbf{Gender}        &   &  &   &\\
Male      &   5 (0.45)&  1 (0.14) & 14 (0.35)  & 20 (0.34)\\
Female    &   6 (0.55)&  6 (0.86)& 26 (0.65)  & 38 (0.66)\\
{\it Sum} &   11 (1.00)& 7 (1.00) & 40 (1.00)  & 58 (1.00)\\\hline
\textbf{Age}        &   &  &   &\\
Under 18 years old  & 1 (0.09)  & -        &  7 (0.18)  & 8 (0.14)\\
18--24 years old    & 7 (0.64)  & 5 (0.71) &  26 (0.65) & 38 (0.65)\\
25--34 years old    & 2 (0.18)  & 2 (0.29) &  5  (0.13) & 9 (0.16)\\
35--44 years old    & 1 (0.09)  &  -       &  2  (0.05) &  3 (0.05)\\
45 years or older   & -         &  -       &  -         & -\\
{\it Sum}         & 11 (1.00) &  7 (1.00) &  40 (1.01)& 58 (1.00)\\\hline
\textbf{Education}        &   &  &   &\\
Basic education           &  -        & -        & 2  (0.05) & 2  (0.05) \\
High school &  6 (0.55) & 4 (0.57) & 17 (0.43) & 27 (0.47)\\
Undergraduate degree         &  5 (0.45) & 3 (0.43) & 19 (0.48) & 27 (0.47)\\
Master's degree           &  -        & -        & 2 (0.05)  & 2 (0.03)\\
Doctoral degree           &  -        & -        &  -        & - \\
{\it Sum}               & 11 (1.00) &  7 (1.00)&  40 (1.01)& 58 (1.00)\\\hline
\textbf{Belief}        &   &  &   &\\
Guaranteed by law &  5 (0.45) & 3 (0.43) & 27 (0.68) & 35 (0.60)\\
Personal responsibility &  6 (0.55) & 4 (0.57) & 13 (0.32) & 23 (0.40)\\
Needed in wrongdoing & - & - & -& -\\
{\it Sum}                            & 11 (1.00)  & 7 (1.00) & 40 (1.00)& 58 (1.00)\\\hline
    \end{tabular}
    \caption{Statistical summary of users' demographics and privacy beliefs in the Pulso system.}
    \label{table:demographicsPulso}
\end{table}

\begin{table}[ht]
    \small
    \begin{tabular}{p{2.503cm}|p{1.11cm}|p{1.11cm}|p{1.0cm}|p{1.0cm}}
    \hline
        & Unconcern. & Fundament. &  Pragmat. & {\it}All \\ \hline
\textbf{Gender}        &           &           &   &\\
Male      & 5 (1.00)  &  10 (0.90)& 35 (0.90)  &  50 (0.90)\\
Female    & -         &  1 (0.10) &  4 (0.10)  & 5 (0.10)\\
{\it Sum} & 5 (1.00)  &  11 (1.00)& 39 (1.00)  & 55 (1.00)\\\hline
\textbf{Age}            &   &  &   &\\
Under 18 years old  & -        & -         & -         &  - \\
18--24 years old    & 1 (0.20) & 2 (0.18)  & 14 (0.36) & 17 (0.31)\\
25--34 years old    & 3 (0.60) & 9 (0.82)  & 24 (0.62) & 36 (0.65)\\
35--44 years old    & 1 (0.20) & -         & 1 (0.03)  & 2 (0.04)\\
45 years or older   & -        & -         & -         & - \\
{\it Sum}         & 5 (1.00) & 11 (1.00) & 39 (1.01) & 55 (1.00) \\\hline
\textbf{Education}        &   &  &   &\\
Basic education           & -        & -        & -         &  -   \\
High school & -        & -        & 4  (0.10) & 4 (0.07)   \\
Undergraduate degree         & 4 (0.80) & 10 (0.91)& 31 (0.79) & 45 (0.82)   \\
Master's degree           & 1 (0.20) & 1 (0.09) & 4 (0.10)  & 6 (0.10)   \\
Doctoral degree           & -        & -        & -         & -   \\
{\it Sum}               & 5 (1.00) & 11 (1.00)& 39 (0.99) & 55 (1.0)   \\\hline
\textbf{Belief}        &   &  &   &\\
Guaranteed by law  & 1 (0.20)  & 5 (0.45) & 28 (0.72) & 34  (0.62)\\
Personal responsibility  & 4 (0.80)  & 6 (0.55) & 11 (0.28) & 21  (0.38)\\
Needed in wrongdoing  &- & - & - & - \\
{\it Sum}         & 5 (1.00) & 11 (1.00) & 39 (1.00) & 55 (1.00) \\\hline
    \end{tabular}
    \caption{Statistical summary of users' demographics and privacy beliefs in the Lumen system.}
    \label{table:demographicsLumen}
\end{table}

These results for each profile reproduce the demographics distribution of the whole sample of users of the systems. Users who exhibit a pragmatist privacy attitude profile stand out in the analysis. {\it They are the majority of users}. Also, differently from users who exhibit the other profiles, {\it users who exhibit a pragmatist privacy attitude profile are more inclined to the belief of privacy as a right guaranteed by law}.

\begin{figure*}[htb]
\centering
\subfigure[Pulso]{
\includegraphics[width=.47\linewidth]{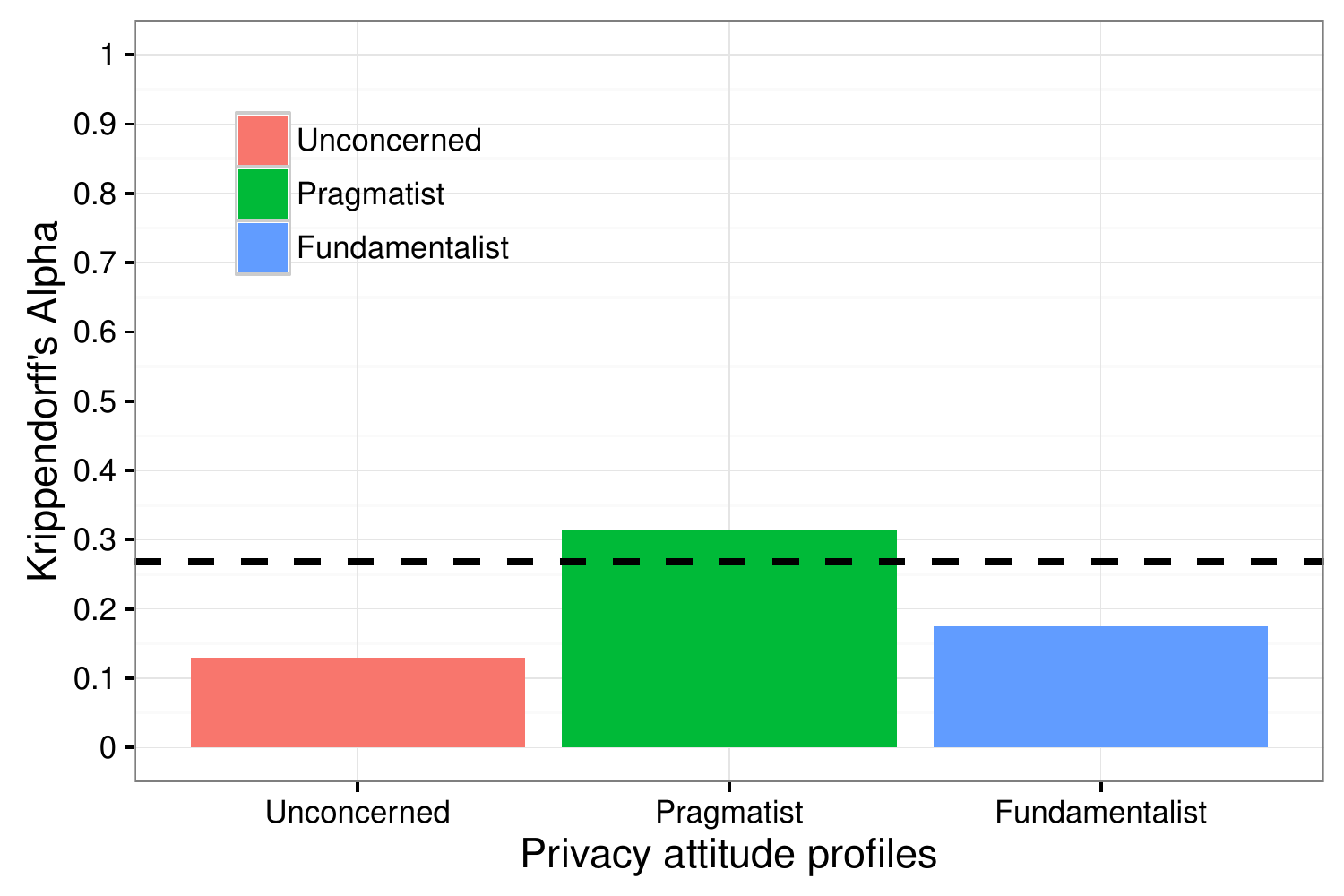}
\label{subfig:pulso:profilesdistribution2}
}
\subfigure[Lumen]{
\includegraphics[width=.47\linewidth]{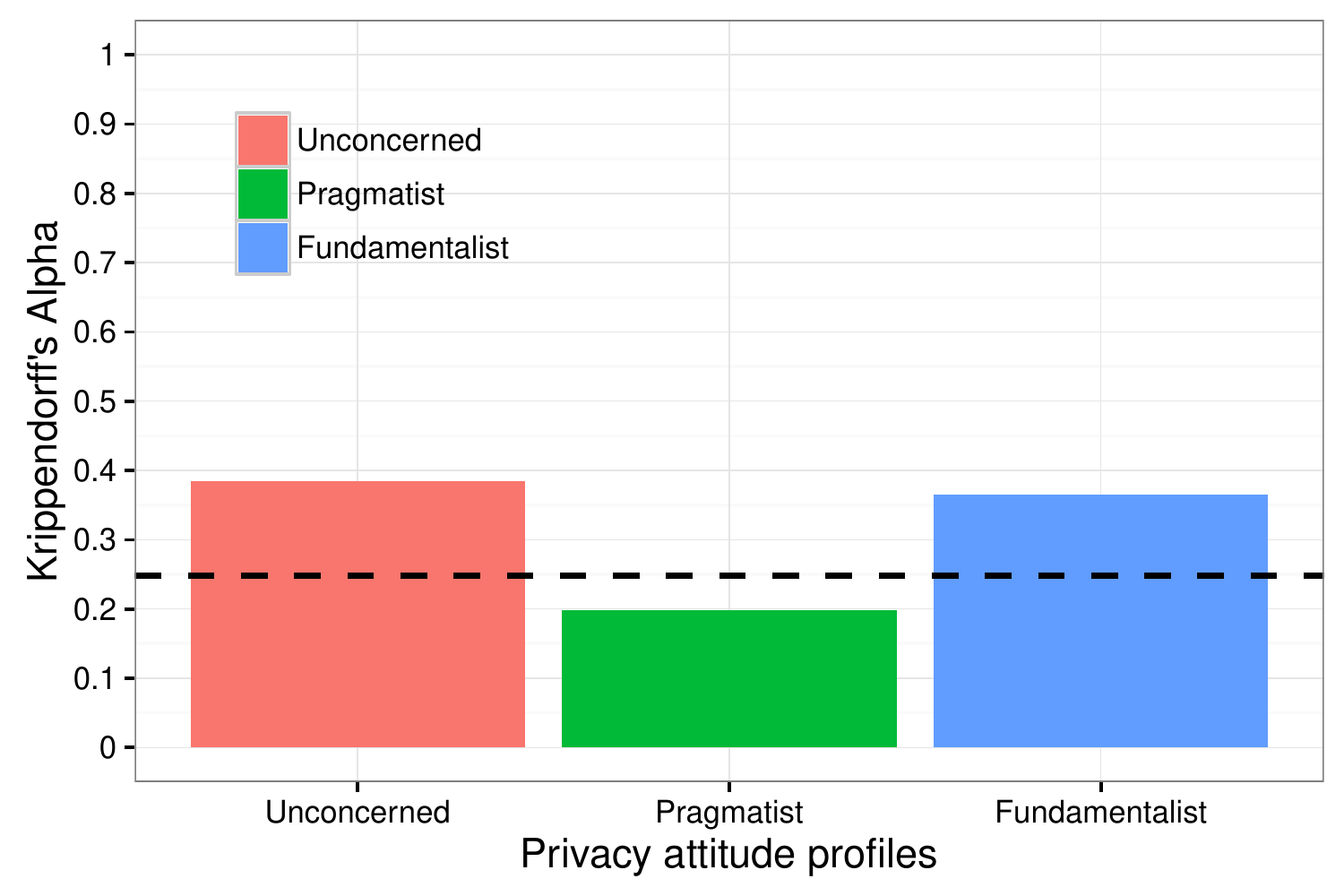}
\label{subfig:lumen:profilesdistribution2}
}
\vspace{-8pt}
\caption{Krippendorff's Alpha consensus coefficient according to users' privacy attitudes profiles. The dashed line is the level of consensus when all users are put together.}
\label{fig:profilesdistribution2}
\end{figure*}

We analyze the level of consensus observed in the answers provided by users to questions about privacy. The result is shown in Figure~\ref{fig:profilesdistribution2}. By putting together the individuals who exhibit the same privacy attitude profiles, {\it we identify groups of individuals who exhibit a level of consensus among them higher than the level of consensus that exists among participants from all profiles}. This is exemplified by the pragmatist profile in the case of Pulso system and the unconcerned and fundamentalist profiles in case of Lumen system. As these results show, the level of consensus among the users about privacy issues is not something that remains invariant across systems.

\subsubsection{The sources of privacy concerns: Collection, inference and exchange of private information}

\begin{figure*}[ht]
\centering
\subfigure[Pulso]{
\includegraphics[width=.47\linewidth]{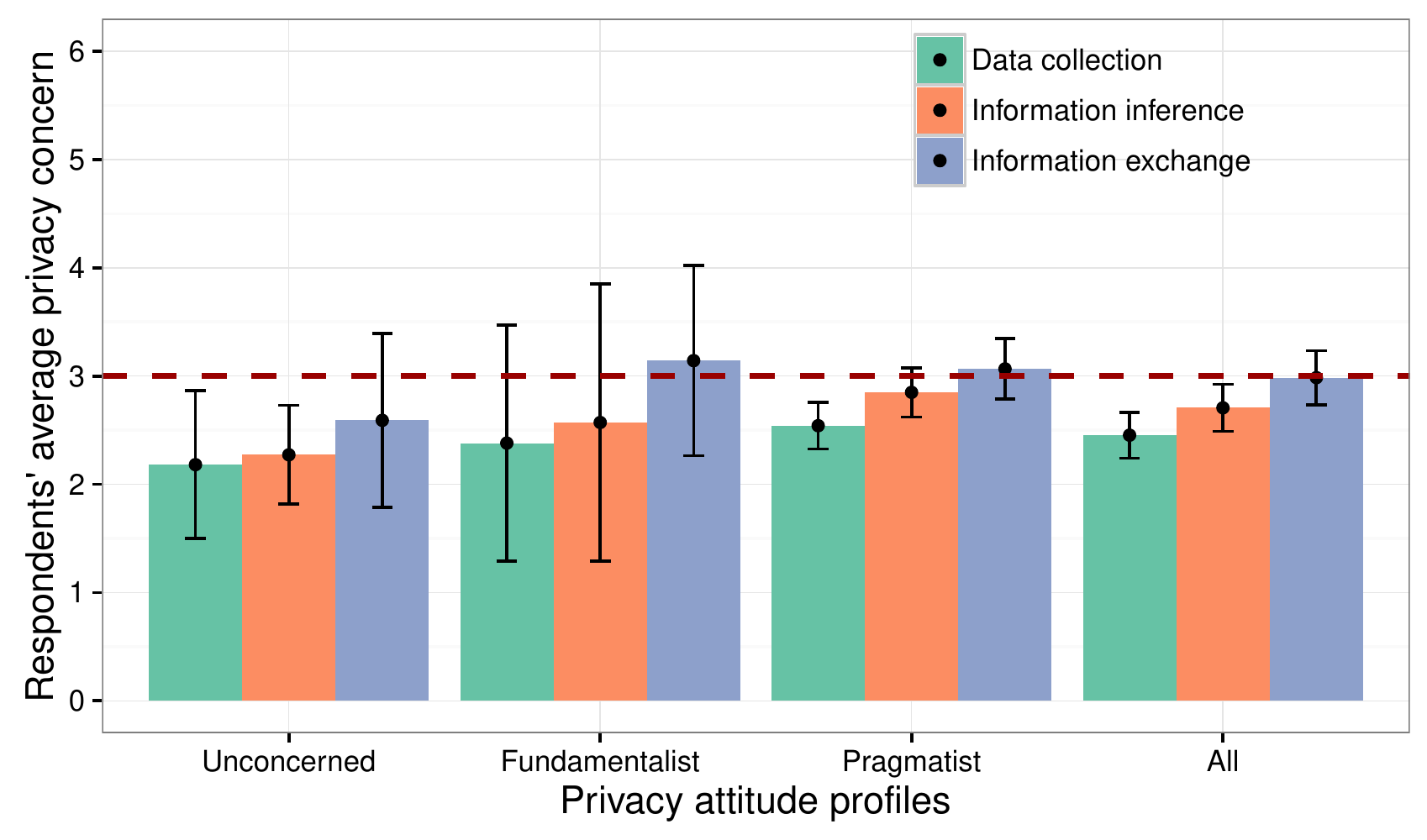}
\label{subfig:pulso:SourcesOfConcern}
}
\subfigure[Lumen]{
\includegraphics[width=.47\linewidth]{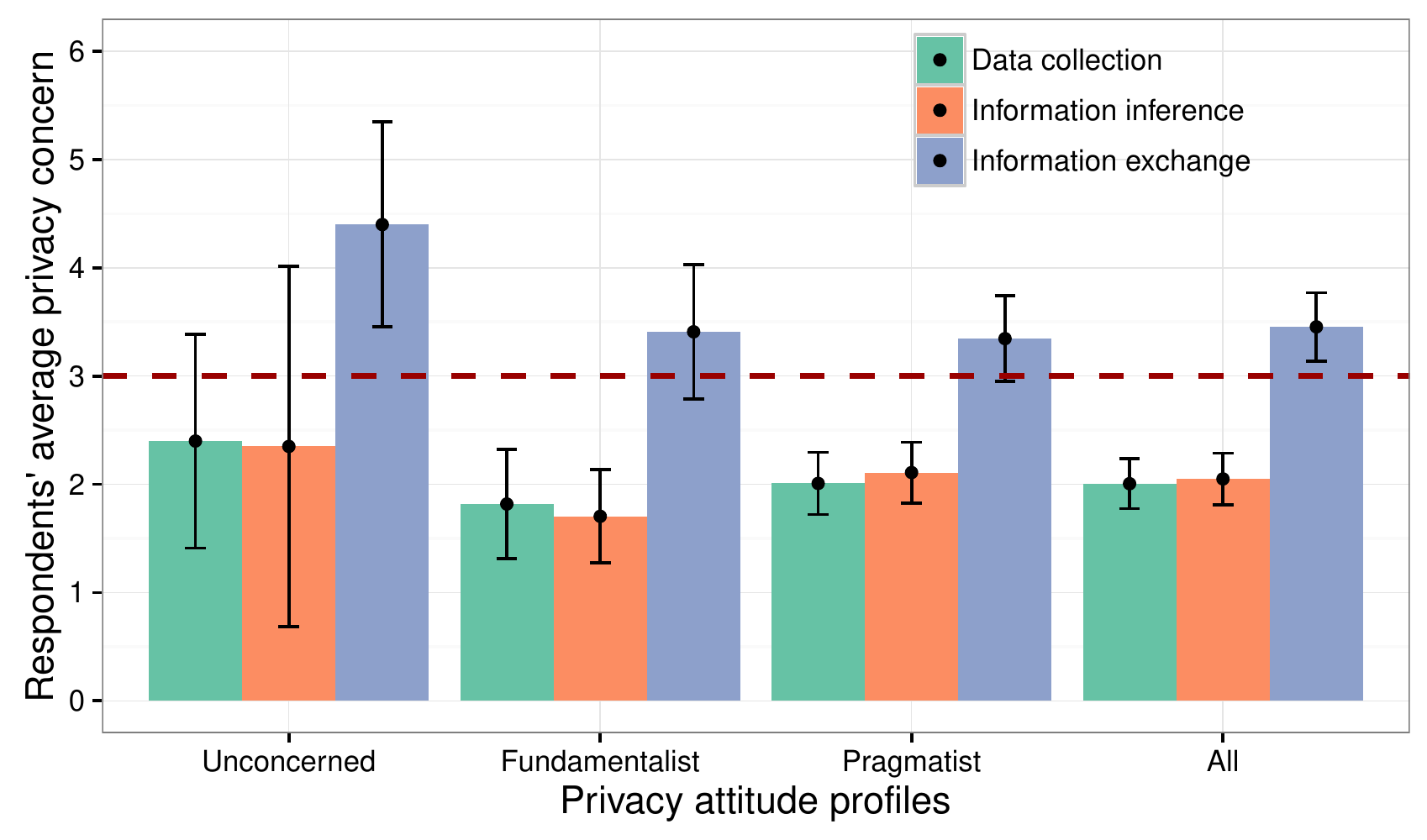}
\label{subfig:lumen:SourcesOfConcern}
}
\caption{Users' average privacy concern about data collection, information inference and information exchange in the system: (a) Lumen and (b) Pulso. The dashed line is the level of privacy concern equivalent to ``Moderately concerned.'' Each error bar represents the 95\% confidence interval.}
\label{fig:sourcesOfConcern}
\end{figure*}

We assess the average level of privacy concern of participants considering the collection of personal data, inference of richer information and exchange of private information with third parties. Figure~\ref{fig:sourcesOfConcern} shows the results according to user's privacy attitude profiles. There are high variations around the average level of concern, mainly in the level reported by privacy unconcerned and fundamentalist users. The results also show a tendency of users to exhibit a higher level of concern about information exchange, sometimes the average lies above the level of privacy concern equivalent to ``Moderately concerned.'' This tendency is observed in all privacy attitude profiles and it is stronger in the case of Lumen system. It indicates that {\it users do not perceive so much problem in the data collection and information inference carried by the system, but they worry that this information will be available to third parties}.   

\begin{figure*}[bth]
\centering
\subfigure[Pulso]{
\includegraphics[width=.47\linewidth]{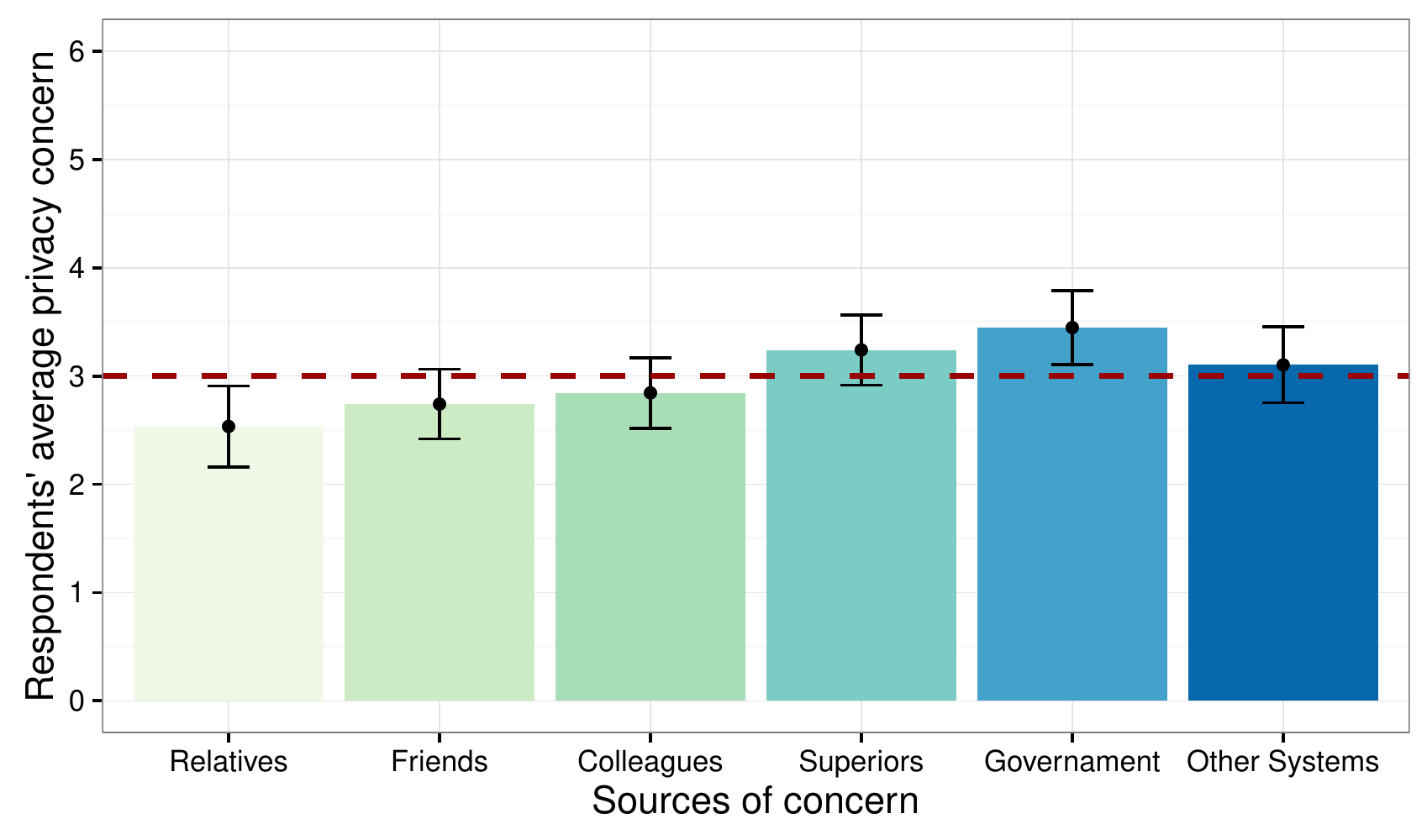}
\label{subfig:pulso:SourcesOfConcernExchange}
}
\subfigure[Lumen]{
\includegraphics[width=.47\linewidth]{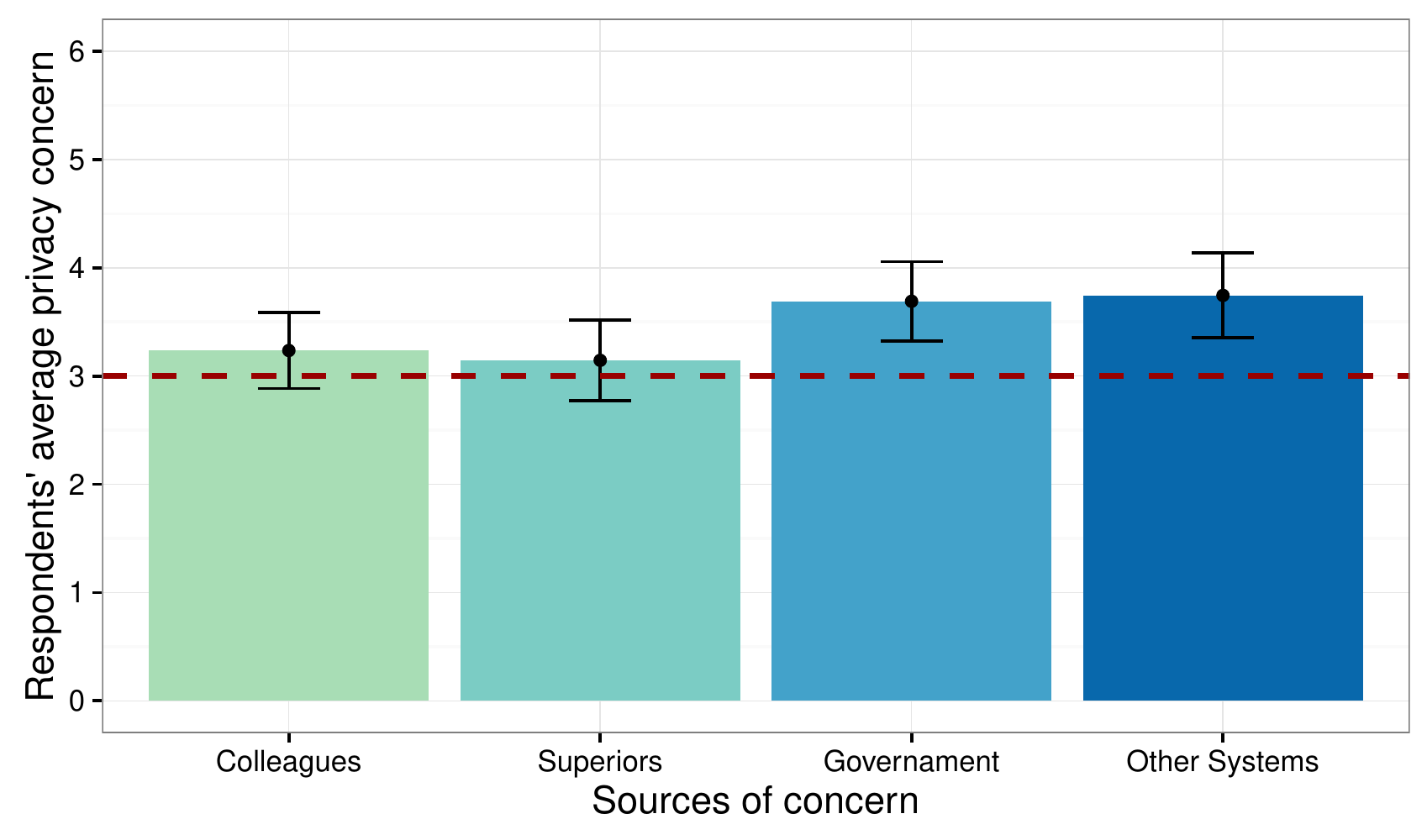}
\label{subfig:lumen:SourcesOfConcernExchange}
}
\caption{Users' average privacy concern about information exchange. The dashed line is the level of privacy concern equivalent to ``Moderately concerned''. Each error bar represents the 95\% confidence interval.}
\label{fig:SourcesOfConcernExchange}
\end{figure*}

To complement our understanding about this result, we analyze which third party users fear most in terms of privacy. As shown in Figure~\ref{fig:SourcesOfConcernExchange}, the difference of concern regarding the investigated sources is small. In both case studies, the only {\it clearly distinguishable high source of concern is government agencies}. Users are concerned about their data being accessible to government agencies. The level of concern lies above the moderated level in both systems. 

\subsubsection{Perceptions on system's data use trade-offs}

We investigate the association between the level of utility that users perceive in a feature and the risk in terms of level of concern they feel about the personal data used by the feature.

In the case of the Pulso system, the feature consists of showing the name and photo of people who are in a specific location. The utility is having access to this information provided by other users and the risk is the level of privacy concern in providing this information to be available to other users. Table~\ref{table:Correlations} shows the correlation between risk and utility in the Lumen system. In the whole set of users, the Pearson correlation between utility and risk is low and not significant. Analyzing the correlation per profile, {\it we find that the fundamentalist users show a significant strong negative correlation}. A simple linear regression was calculated to model this relationship between utility and risk exhibited by the fundamentalist users. A significant regression equation was found (F(1,5)=6.80, p-value$<$0.05), with an adjusted $r^2$ of 0.49. In this case, the risk is 4.82-0.75(utility). The risk perceived by users decreases 0.75 to each unit increased in utility.

\begin{table}[ht]
\small
\centering
\begin{threeparttable}
  \begin{tabular}{ l | l | l }
    \hline
    Privacy attitude profiles & Pulso system & Lumen system\\ \hline
    Unconcerned & -0.15 & -0.37 \\
    Fundamentalists & -0.76* & -0.45 \\
    Pragmatist & ~0.03 & -0.57** \\
    All together & -0.16 & -0.50** \\
    \hline
  \end{tabular}
  \begin{tablenotes}
      \footnotesize
      \item Significance codes:  *p-value $<$ 0.05, **p-value$<$0.001.
    \end{tablenotes}
  \caption{Correlation between risk and utility according to privacy attitude profiles in the Pulso and Lumen systems.}
  \label{table:Correlations}
\end{threeparttable}
\end{table}

In the case of the Lumen system, the studied feature consists of showing the name of people who are related to an energy consumption problem in the organization. The utility is the usefulness of this feature and the risk is having his/her the name and photo exposed in this situation. In the whole set of users, the Pearson correlation between the level of perceived utility and the level of perceived concern is moderate and significant. Analyzing the correlation per profile, {\it only pragmatist users show a significant correlation}, the unconcerned and fundamentalist users exhibit no significant correlations. To the whole set of users, a significant regression equation was found (F(1,53)=17.42, p-value$<$0.001), with an adjusted $r^2$ of 0.23. In this case, the risk is 4.74-0.52(utility). Users' perceived risk decreases 0.52 to each unit of utility. A significant regression equation was also found to the pragmatist users (F(1,37)=13.55, p-value$<$0.001), with an adjusted $r^2$ of 0.24. In this case, the risk is 4.75-0.52(utility). The risk perceived by the users decreases in 0.52 to each unit increased in utility.

In general, these results indicate a negative relation between utility and risk. This relation means that people tend to be more concerned about the provision of data to features that they do not perceive usefulness or, alternatively, they tend to penalize the level of utility of features that make use of information that causes them much privacy concern.

\section{Discussions and Design Implications}
\label{sec:discussions}

Our research helps to answer the question about the underlying factors of users' perceptions in IoT systems. It emphasizes users' privacy beliefs, attitudes, systems' source of concern, and risk-utility trade-offs. The statistical analysis showed a diversity of privacy perception and differences among the individuals. In the following, we first discuss these results, then we propose a set of heuristics to cope with privacy concerns in the design of IoT systems.

\subsection{Privacy in IoT systems}

Our analysis allowed us to identify users who exhibit the three privacy attitude profiles: unconcerned, fundamentalists, and pragmatist. We show that, as in other contexts, most users are pragmatists. Our results suggests that users who exhibit a pragmatist profile tend to believe in privacy as a right guaranteed by law, while users who exhibit unconcerned and fundamentalists profiles balance their beliefs between privacy as a right and privacy as a personal responsibility.

Privacy is a subject in which people often have different opinions and low consensus among them because it relies on several personal characteristics. Our results show that in IoT systems even users who fall in the same privacy attitude profile can diverge among them in terms of their privacy beliefs and concern. Users' privacy attitude profiles have an effect on the level of consensus obtained in the responses provided by them to privacy issues. In our analysis, the level of consensus about privacy issues is not something that remains invariant across systems.

The major source of privacy concerns in IoT systems tends to be the exchange of user personal data to third parties. Depending on the context of the system, the level of the privacy concern caused by data exchange exceeds the moderated level. The third party that caused the higher levels of concern is the government, but other third parties as people hierarchically superiors and other systems can also cause high levels of concern depending on the context.

Our analysis of risk-utility trade-off indicates a negative correlation between utility and risk. The higher the perception of utility, the lower the perception of risk. This correlation means that users tend to penalize the level of utility of features that may cause privacy concerns.

Finally, we believe that some level of ``concern about privacy'' is an unavoidable consequence of sharing personal information. Even users who exhibit an unconcerned privacy attitude profile tend to exhibit some level of concerns, though low. Concerns occur in people's daily lives and are natural that they also occur in their interaction with IoT systems. The design of the system should focus on reducing the concerns to levels that are comfortable to the users.

\subsection{Heuristics to cope with privacy concerns in IoT}

To summarize the insights gained from our analysis of users' privacy perceptions and attitudes, we present design heuristics that can be expected to support the design and configuration of IoT systems following a ``design for privacy'' perceptive~\cite{Bellotti:1993}.

\subsubsection{Let users know what information the system has about them.} Users may not properly understand the data collection process because data can be collected indirectly via their devices and such collection is intrinsically ubiquitous. One way for the users to know what kind of data is collected and inferred about them is to allow the user to see and download that data.

\subsubsection{Make clear the usefulness of the data for each feature.} Users tend to be less concerned about privacy when they understand the utility of the feature that make use of the collected data. If the utility of the feature is low, any data collected or inferred by the system will let users feel high levels of concern.

\subsubsection{Make the exchange of data with third parties explicit and configurable.} The provision of users data to third parties is one of the major sources of privacy concern. This can not be done without the knowledge and the explicit authorization of the user, so it can not be a default system configuration. System should also give to users the possibility to specify to which third party their data can be provided, if any. It allows one to address the variances in the level of concern that exists in the set of users.

\subsubsection{Conduct empirical assessments of privacy.} Our results show several privacy issues that cannot be generalized across systems. New systems may be based on features that manifest significant privacy risks to their users. The effects of such features on users' privacy should be tested before the system is made available to its users.

\subsection{Threats to the validity}

This study was conducted in Brazil and had students and technology professionals as participants. The results are based on answers provided by the participants, which is subject to the social desirability effect. Responses provided by participants may differ from their behavior when using the system; which is known as privacy paradox~\cite{barnes2006privacy,norberg2007privacy,Williams2016}.

While the proposed heuristics will be helpful in ensuring the presence of basics privacy requirements, it should not lead designers to assume that they are enough to design privacy-friendly systems. We are convinced that such heuristics interplay with other initiatives, such as security mechanisms, compliance with legislation and good data governance practices. 

\section{Conclusions}
\label{sec:conclusions}

In this study, we sought to deeply investigate individuals privacy perceptions about systems based on the Internet of Things. Building on previous studies, we proposed a survey approach to characterize users' perceptions. The characterization focuses on individuals privacy beliefs, components of IoT systems that cause privacy concern (data collection, information inference, and information exchange), and perceived risk-benefit trade-offs in the features provided by the system. 

We survey 113 individuals about their privacy perceptions in two IoT systems: Lumen and Pulso. We classify the individuals into three privacy attitude profiles: unconcerned, fundamentalists and pragmatists. We found that most of the individuals exhibit a pragmatist attitude towards privacy and tend to believe in privacy as a right guaranteed by law. The exchange of information to third parties tends to be the most concerning aspect compared to data collection and information inference. The perceived privacy risk tends to be negatively associated with the utility; the higher the perception of utility, the lower the perceived risk. We derive design heuristics to help system designers to cope with user privacy concerns.

As future work, we suggest the extension of this study in order to also investigate the effects of national, cultural and economic factors. We also suggest the conduction of an observational study with users over their interaction of the systems. Besides of how privacy rules (concerns and attitudes) are formed in IoT system, such study would also characterize how the rules change over time.

\section{Acknowledgments}
This research was partially funded by EU-BRA SecureCloud project (MCTI/RNP 3rd Coordinated Call) and by CNPq, Brazil.

\balance{}

\bibliographystyle{SIGCHI-Reference-Format}
\bibliography{sample}

\end{document}